\documentstyle[12pt]{article}
\begin{document}
\begin{titlepage}
\pagestyle{empty}
\baselineskip=21pt
\rightline{\small hep-th/yymmddd \hfill McGill 97--27}
\vskip 1.5in
\begin{center}
{\huge{\bf Wavy Horizons?}}
\vskip 1.0in
 
{\large
Robert C. Myers\footnote{Based on a talk presented at
the Seventh Canadian Conference on General Relativity
and Relativistic Astrophysics in Calgary, AB, Canada,
May 31 -- June 7, 1997.\\
E-mail: rcm@hep.physics.mcgill.ca}}
\vskip 1.5em
 {{\it Department of Physics, McGill University,
Montr\'eal, PQ, H3A 2T8 Canada}}
\vskip 4em
 
\begin{abstract}
We describe the application of a gravity wave-generating technique to certain
higher dimensional black holes. We find that the induced waves generically
destroy the event horizon producing parallelly propagated curvature
singularities.
\end{abstract}
\end{center}

\end{titlepage}
\baselineskip=18pt
 
\setcounter{footnote}{0}

\def\id{1\!\!1}
\def\beq{\begin{equation}}
\def\eeq{\end{equation}}
\newcommand{\beqa}{\begin{eqnarray}}
\newcommand{\eeqa}{\end{eqnarray}}
\newcommand{\beqar}{\begin{eqnarray*}}
\newcommand{\eeqar}{\end{eqnarray*}}
\renewcommand{\a}{\alpha}
\newcommand{\al}{\alpha}
\renewcommand{\b}{\beta}
\newcommand{\del}{\delta}
\newcommand{\D}{\Delta}
\newcommand{\eps}{\epsilon}
\newcommand{\g}{\gamma}
\newcommand{\G}{\Gamma}
\renewcommand{\l}{\lambda}
\renewcommand{\L}{\Lambda}
\newcommand{\na}{\nabla}
\renewcommand{\O}{\Omega}
\newcommand{\p}{\phi}
\newcommand{\s}{\sigma}
\renewcommand{\t}{\theta}
\newcommand{\AH}{S}
\newcommand{\eg}{{\it e.g.,}\ }
\newcommand{\ie}{{\it i.e.,}\ }
\newcommand{\pol}{\varepsilon}
\newcommand{\labell}[1]{\label{#1}} 
\newcommand{\labels}[1]{\label{#1}} 
\newcommand{\F}[1]{{[F_{(#1)}]}} 
\newcommand{\T}[1]{{[\theta_{(#1)}]}} 
\newcommand{\Fp}[1]{{[F'_{(#1)}]}} 
\newcommand{\B}{B_0}
\newcommand\ee{{\rm e}}
\renewcommand\d{{\rm d}}
\newcommand\reef[1]{(\ref{#1})}
\newcommand\prt{\partial}

\section{INTRODUCTION}

Interest in black hole uniqueness theorems began some thirty years ago
with the pioneering work of Israel\cite{firsthair}. The no-hair
results are now rigorously established for Einstein gravity coupled to
Maxwell fields and various other simple matter systems\cite{hair}.
More recently, physicists have become interested in theories
with more complicated matter field couplings, as well
as spacetime dimensions beyond four. While this research has
produced a plethora of new solutions\cite{newsols}, they are found to
respect the spirit of the no-hair theorems in that
the black hole geometries are still
completely determined by some small set of charges.

An interesting corollary of the early theoretical investigations
of black holes was that each connected component
of a stationary horizon must have the topology of a two-sphere\cite{round}.
One might regard this result as indicating black holes carry 
no `topological hair.' This result is easily evaded, however, in higher
dimensions. As a simple example, consider the four-dimensional
Schwarschild metric combined with a flat metric on $R^n$. This space-time
is an extended black hole solution of Einstein's 
equations in 4+$n$ dimensions, and the topology of the horizon is $S^2\times
R^n$. Clearly, this straightforward construction is easily extended to
constructing
many other higher-dimensional black holes whose horizons inherit the topology
of the `appended' manifold.\footnote{Similar solutions arise for four
dimensions in the presence of a negative cosmological constant\cite{neg}.}

While higher-dimensional black holes might have `topological hair', 
one still expects that the spirit of the no-hair theorems should
be obeyed in these cases. So having
fixed the horizon topology and a limited set of charges, the black hole
solution should be completed determined. Our present
work focuses on a potential violation of these expectations.
For certain classes of solutions, it is possible to generate
an infinite variety of new solutions by adding
wave-like perturbations\cite{GV}. These techniques may be
applied to certain extended black objects, and would apparently
yield black holes with an infinite variety of wavy hair.
The paper is organized as follows:
Section~\ref{sec:solution} describes the wave-generating technique, and
provides a simple example of a wavy black hole. 
Section~\ref{sec:singularity} investigates the curvatures
of these wavy `horizons' and section~\ref{sec:end}
provides a brief discussion of our results. While the present
discussion is self-contained, it is lacking in many details
which the interested reader may find in ref.~\cite{us}.

\section{WAVES \labels{sec:solution}}

We begin with a brief review of the wave-generating technique of Garfinkle
and Vachaspati\cite{GV}. 
This method was originally developed in the context of the Yang-Mills-Higgs
system coupled to gravity. However, it is straightforward to
extend the construction to general supergravity (or low-energy
string) theories in arbitrary dimensions\cite{us}. 
The starting point for this construction is a solution
with a vector field $k^\mu$ which is
null, hypersurface orthogonal and Killing, \ie
\beq
\labell{rules} 
k^\mu k_\mu=0\ ,
\qquad\quad
\nabla_{\![\mu} k_{\nu]} = k_{[\mu}\nabla_{\!\nu]}\AH\ ,
\qquad\quad
\nabla_{\!(\mu} k_{\nu)} = 0\ .
\eeq
If the solution contains non-trivial matter fields,
their Lie-derivative must also vanish
--- and certain transversality constraints\cite{us} must be
satisfied as well ---
in order that $k^\mu$ yields an invariance of the full solution.
One then defines a new metric\cite{GV}
\begin{equation} \labell{w31}
\tilde{g}_{\mu\nu} = {g}_{\mu\nu} + \ee^\AH\, \Psi\, k_{\mu}\,k_{\nu},
\end{equation}
while leaving all of the matter fields unchanged.
The new metric $\tilde{g}_{\mu\nu}$ will also be a solution provided that
the function $\Psi$ satisfies appropriate constraints. 

The first constraint is that $k^\mu$ has a vanishing Lie-derivative
on $\Psi$. One may verify that this also ensures that
the hypersurface orthogonal and Killing conditions 
are still satisfied with the new metric
(and with the same $\AH$).\footnote{Note it follows directly for
eq.~(\ref{rules}) that $k^\mu$ has a vanishing Lie-derivative on $\AH$.}
It is also obvious that $k^\mu$ remains null 
with $\tilde{g}_{\mu\nu}$. The remaining restrictions arise
to guarantee that after the metric is shifted,
the explicit form of the equations of motion remains unchanged.
One finds that the matter field equations
are automatically invariant. However, invariance of 
Einstein's equations requires 
that the d'Alembertian acting on $\Psi$ vanish.
Therefore given a solution with a vector $k^\mu$ satisfying eq.~\reef{rules},
eq.~(\ref{w31}) yields a new solution provided
\beq\labell{psiconstr}
k^\mu\prt_\mu\Psi=0 \qquad{\rm and}
\qquad \nabla^2 \Psi=0\ \ .
\eeq

These perturbations can be interpreted
as gravity waves as follows: Consider a coordinate system adapted to
the flow of $k^\mu$. As well as the cyclic coordinate $v$,
there is a coordinate $u$ given `roughly' by the integral of the
dual one-form $k=k_\mu dx^\mu$. As none of the fields depend on 
the Killing coordinate $v$, the only `time' dependence can
arise through this null coordinate $u$. Therefore, the
perturbations are moving through the space-time
at the speed of light along the $u$ direction.

While eq.~\reef{rules} is very restrictive, this solution-generating
technique has found a wide-range of applications\cite{GV,rest,dgr,HM}.
We will present a particular black hole which arises in
low-energy superstring theory satisfying these symmetry restrictions
\reef{rules}:
\beqa
ds^2&=&- \left(1-{Q\over R}\right)^2 dt^2+
{dR^2\over\left(1-{Q\over R}\right)^2}+ R^2(d\theta^2
+\sin^2\theta\,d\phi^2)
\nonumber\\
\labell{sol11}
&&\qquad\qquad+\left(dy-{Q\over R}dt\right)^2
+\sum_{i=5}^9 (dx^i)^2\ .
\eeqa
Within ten-dimensional superstring theory, the complete solution includes
various matter fields whose details are inessential to the
following discussion. Amongst the higher dimensions,
$y$ is distinguished by the nonvanishing $G_{yt}$ component,
which indicates that momentum is flowing in this
particular direction. The reader may recognize the first four terms
in the line element as describing the geometry of the extremal
Reissner-Nordstrom black hole. This is the four-dimensional
space-time which would be observed by low-energy physicists. 
The full solution then describes a black hole with a horizon
at $R=Q$ which has the topology of $S^2\times R^6$. 

As is evident from eq.~\reef{sol11}, this
solution has a number of Killing vectors and it is straightforward
to show that the combination
\begin{equation}
\labell{sol23}
k^\mu\,\partial_\mu\equiv\partial_v=\partial_t+\partial_y
\end{equation}
is everywhere null. In fact, one finds that $k^\mu$ is the null generator
of the horizon.
The $\partial_{y}$ contribution then indicates linear motion
of the horizon along the $y$ direction. From 
\beq
k_\mu dx^\mu=\left(1-{Q\over R}\right)(dy-dt)
\eeq
one sees that 
the null Killing vector $k^\mu$ is also hypersurface-orthogonal,
with $\ee^{-S}=1-Q/R$.
Hence the metric \reef{sol11} admits the symmetry \reef{rules} required
to apply the wave-generating technique. 

Using eq.~(\ref{w31}), one constructs a new metric
\begin{equation}
\labell{sol25}
\widetilde{ds}^2=ds^2+\ee^{-S}\,\Psi\,(dy-dt)^2
\end{equation}
where $\Psi$ must satisfy the constraints in eq.~(\ref{psiconstr}).
The first condition $(\partial_t+\partial_{y})\Psi=0$ requires
$\Psi=\Psi(u\!=\!t\!-\!y,r,\theta,\phi,x^i)$. Before examining the second
constraint, it is convenient to shift the
radial coordinate to $r=R-Q$, which transforms eq.~\reef{sol11} to
\beqa
ds^2&=&- f^{-2}{dt^2}+
f^2\left(dr^2+ r^2(d\theta^2+\sin^2\theta\,d\phi^2)\right)
\nonumber\\
\labell{solman}
&&\quad+\left(dy-{Q\over fr}dt\right)^2
+\sum_{i=5}^9 (dx^i)^2
\eeqa
with $f\equiv 1+Q/r$.
With this coordinate shift, the second condition reduces to
\beq
\labell{soulman}
\left[ \nabla^2_F+ f^2\sum_{i=5}^9(\partial_i)^2\right]\Psi=0
\eeq
where $\nabla^2_F$ is the Laplacian on the {\it flat}
space covered by ($r,\theta,\phi$). The Killing constraint
has ensured that no $t$ or $y$ derivatives appear in
eq.~\reef{soulman}.

For simplicity, we begin by considering solutions of eq.~\reef{soulman}
which are independent of the internal coordinates $x^i$.
In this case, the general solution is
\beq\labell{Psol}
\Psi=\sum_{l,m}\left(a_{lm}(u)\,r^l+b_{lm}(u)\,r^{-(l+1)}\right)
P^m_{l}(\cos\theta)\cos(m\phi+\delta_m(u)\,)
\eeq
where $u=t-y$, as above, and the
$P^m_{l}(\cos\theta)$ are associated Legendre functions.
In general, the phases $\delta_m$, as well as the amplitudes
$a_{lm}$ and $b_{lm}$, are {\it arbitrary} functions of $u$.
Now there are two classes of solutions, those which
grow at large $r$ and those which decay. 
In the former case with $r^l$ and $l>1$,
the metric is not asymptotically flat\footnote{In fact,
the $r^l$ perturbation with
$l=0$ yields a diffeomorphism of the original metric, while
with $l=1$ the solution is asymptotically flat and the perturbation
produces transverse oscillations of the horizon \cite{dgr,CMP2}.},
and so these
perturbations are not intrinsic to the black hole, rather
they fill the asymptotic region with gravitational radiation.
Hence I will focus on the decaying solutions
with $r^{-(l+1)}$ profiles.
These perturbations are localized near the `horizon' at $r=0$,
and hence are candidates for `wavy' hair on the black hole.

Generalizing to solutions of the full equation \reef{soulman}, one
finds 
\beq\labell{Psolb}
\Psi=\sum_{l,m,n_i}b_{lmn_i}P^m_{l}(\cos\theta)\cos(m\phi+\delta_m)
\prod_{i=5}^9
\cos\left(n_ix^i/R_i+\delta_{n_i}\right)\,F_{ln_i}(r)
\eeq
where the $x^i$ coordinates are assumed to be compactified
with period $2\pi R_i$. Again, the phases $\delta_m$ and $\delta_{n_i}$
and the amplitudes $b_{lmn_i}$ can have an arbitrary dependence on $u$.
The radial functions $F_{ln_i}$ satisfy
\beq
\labell{radeq}
\left[{\partial^2\ \over\partial r^2}+{2\over r}{\partial\ \over\partial r}
-{l(l+1)\over r^2} -\left(1+{Q\over r}\right)^2 M^2\right]\,F(r)=0
\eeq
where $M^2=\sum_{i=5}^9(n_i/R_i)^2$. The solutions can be written in
terms of confluent hypergeometric functions, but only a
qualitative description of the solutions will be needed here.
In general, there are again
two classes of solutions: growing and localized.
As before, the growing perturbations will be discarded as they fill
the entire spacetime with gravitational waves.
The localized solutions are more interesting, as they
are candidates for wavy hair. Their long range
behavior is $F\sim\exp(-Mr)/r$, and hence any
internal oscillations result in perturbations which decay faster than any
of those in eq.~\reef{Psol}. In the limit $r\rightarrow0$,
the solutions admit a series representation of the form:
$F=r^{-\beta}(1+\sum_{n=1}^\infty F_n r^n)$.
From eq.~(\ref{radeq}), one finds that the leading power is
\beq
\labell{exponb}
\beta = (1 +\sqrt{1 + 4l(l+1)+4M^2Q^2}\,)/2\ .
\eeq
Hence one finds that all of the candidates for wavy hair 
have singular behavior at the null surface $r=0$.

\section{SINGULARITIES\labels{sec:singularity}}

The above construction appears to have
provided the black hole \reef{solman} with an infinite variety
of wavy hair. Such a result would certainly run contrary to the
idea that higher dimensional black holes should have no hair.
However, the radial profile of these waves diverges as $r^{-\beta}$
near the `horizon,' and hence one must
be careful to investigate whether any curvature singularities
are produced at the surface $r=0$.

A natural approach to investigate this question is to examine various curvature
scalars, \eg $R_{\mu\nu}R^{\mu\nu}$ or $R_{\mu\nu\alpha\beta}
R^{\mu\nu\alpha\beta}$, for the existence of singularities.
Upon performing the lengthy calculations to construct
these scalars, one finds no evidence of a singularity at $r=0$. In fact,
one finds no evidence of the wave perturbation at all! The latter is
true for all curvature scalars, which was proven with
the following theorem\cite{us}:

{\it If  $g_{\mu\nu}$ is a pseudo-Riemannian metric admitting
a null, hypersurface-orthogonal, Killing vector $k^\mu$ and
$\tilde{g}_{\mu\nu}= g_{\mu\nu} + \kappa\, k_{\mu}k_{\nu}$, where
$\kappa$ is any scalar Lie-derived by $k^\mu$ to zero, \ie
${\cal L}_k \kappa = 0$, then all of the scalar curvature invariants of
$\tilde{g}_{\mu\nu}$ are exactly identical to the corresponding
curvature invariants of $g_{\mu\nu}$. }

\noindent This result is purely geometric
in nature, and holds for any metric satisfying the symmetry conditions
in eq.~\reef{rules}. The transformation
(\ref{w31}) provides a specific example where the theorem
applies with $\kappa = \ee^\AH \Psi$.
Hence one has the rather surprising result that all scalar
curvature invariants are identical for both the original and the
shifted metrics in eq.~\reef{w31}. Therefore these invariants
do not contain any information about how the space-time geometry
is modified by the wavy perturbations.
Tidal forces prove to be a better probe of the wave geometry.
These forces are determined by the Riemann curvature measured in the
rest frame of a geodesic observer.

The tidal forces will be calculated for each mode of oscillation
individually. Hence consider as a candidate perturbation:
\beq
\Psi = B(u) F(r)
P^m_l(\cos\theta) \cos \left(m \phi+\delta_m\right)
\prod_{i=5}^9
\cos\left(n_ix^i/R^i+\delta_{n_i}\right)\ .
\labell{constprof}
\eeq
The calculation proceeds in several steps: First, one must show
there exists a geodesic stretching 
between asymptotic infinity and the null surface --- hence showing
that $r=0$ belongs to the space-time.
Next, a convenient orthonormal frame is constructed.
Examining the curvature in this stationary frame, one 
finds that all of the components are finite.
One then constructs the Lorentz transformation relating the
stationary frame to the
rest frame of an observer moving along the geodesics identified
above. Finally, the curvature is boosted to the infalling frame
in order to determine the observer's tidal forces.
The divergences identified in this way are equivalent to
parallelly propagated curvature singularities.

The first step of identifying geodesics 
in the presence of a general oscillation proves to be a daunting 
task. To simplify the problem, we consider the case where
the amplitude $B$ and the phases $\delta$ are constants.
These solutions will be enough to 
identify the leading divergences. The simplest approach 
is to choose values of $\theta,$ $\phi$
and $x^i$ such that the derivatives of $\Psi$
with respect to these coordinates vanish.
Then the geodesic equations are consistently solved with these fixed
values and the motion reduces to
\beqa
dt/d\tau&=& f\,\left[\,\omega +\left({Q/r}+B_0F(r)\right)
(\omega-p)\right]
\nonumber\\
dy/d\tau&=& f\,\left[\,p +\left({Q/r}+B_0F(r)\right)
(\omega-p)\right]
\labell{geod2}\\
dr/d\tau &=& -f^{-1}\left[f^2H^2(\omega-p)^2+{2f}(\omega-p)p
-1\right]^{1/2}
\nonumber
\eeqa
where $\omega$ and $p$ are integration constants, $f\equiv 1+Q/r$
and $H^2\equiv 1+B_0F(r)/f$. We have also set $\Psi=B_0F(r)$ 
along the geodesic at fixed values of $\theta$, $\phi$
and $x^i$. Now in order that the geodesic reaches $r=0$,
the fixed coordinates must be chosen so that $B_0>0$.
Also $\omega^2>p^2+1$ ensures that
the geodesic extends back to $r\rightarrow\infty$. 

As an intermediate step, we define a stationary orthonormal basis
of one-forms
\beqa
e^t &=& dt/fH
\qquad 
e^r = f\,dr
\qquad
e^{y} = H \left(dy-dt\right)  + dt/fH
\nonumber\\
e^\theta &=& (r+Q)\,d\theta
\qquad
e^\phi = (r+Q)\sin\theta\,d\phi
\qquad
e^i=dx^i\ .
\labell{funf}
\eeqa
One can readily verify
that $\widetilde{ds}^2=\eta_{ab}\,e^ae^b$ reproduces the line element
in eq.~\reef{solman}. Calculating the curvature in this frame,
one finds that all components are everywhere finite.
We are particularly interested in the limit
$r\rightarrow0$ along the geodesics identified above.
There the curvature components reduce to
\begin{eqnarray}
R^{trtr}&\simeq&{1-2\beta(\beta-1)\over4Q^2}
\qquad
R^{yryr}\simeq-{1+2\beta(\beta-1)\over4Q^2}
\qquad
R^{tryr}\simeq-{\beta(\beta-1)\over2Q^2}
\nonumber\\
R^{tatb}&\simeq&\delta^{ab}\,l(l+1)/4Q^2\simeq R^{yayb}
\simeq R^{ta yb} \qquad\ {\rm for}\ a,b=\theta,\phi
\nonumber\\
R^{titj}&\simeq&\delta^{ij}{n_i^2/2R_i^2}\simeq R^{yiyj}\simeq
R^{tiyj} \qquad\qquad\quad{\rm for}\ i,j=5,\ldots,9
\nonumber\\
\labell{curvapp}
R^{tyty}&\simeq&{1/4Q^2}\qquad\quad
R^{\theta\phi\theta\phi}\simeq {1/Q^2}\ .
\eeqa
Referred to this frame, the proper ten-velocity is
$V^a = e^a{}_\mu\, dx^\mu/d\tau$:
\beqa
V^t &=&  Hf(\omega-p) +{p}/{H}
~~~~~~~~~~~  V^y = {p}/{H}
\nonumber\\
V^r &=& - \left[f^2H^2(\omega-p)^2+{2f}(\omega-p)p
-1\right]^{1/2}
\labell{4vel}
\eeqa
along with $V^\theta=0=V^\phi=V^i$.
As a check, one may easily verify that $\eta_{ab}V^a V^b=-1$.

Now we need a Lorentz transformation which takes the unit time-like
vector $N^a=\delta^a_t$ into the observer's ten-velocity:
$V^a=L^a{}_bN^b$. Applying this transformation to our stationary
vielbein (\ref{funf}) produces a natural basis of orthonormal
one-forms which the infalling observer might use in her rest frame.
The simplest choice is
\begin{equation}
\labell{loren}
L^a{}_b=~\pmatrix{
V^t & V^y& V^r & 0 \cr
V^y & 1+{(V^y)^2\over V^t+1} & {V^y\,V^r\over V^t+1}& 0 \cr
V^r & {V^y\,V^r\over V^t+1}& 1+{(V^r)^2\over V^t+1} &0 \cr
0 & 0 & 0 & \id_7  \cr}
\eeq
where $\id_7$ is a $7\times7$ identity matrix.

Finally we turn to the tidal forces which the observer
experiences  as $r\rightarrow0$. Here, one boosts
curvature components $R^{klmn}$ calculated in the stationary frame 
to the observer's rest frame with
$\hat R^{abcd} = L^{a}{}_k L^{b}{}_l L^{c}{}_m L^{d}{}_n R^{klmn}$.
Given that all of the components of $R^{klmn}$ are finite,
divergences can only arise through the boost factors.
Hence, we consider the behavior of the ten-velocity \reef{4vel}
as $r\rightarrow0$:
$V^t\simeq (\omega-p)\sqrt{B_0 Q}\, r^{(-\beta-1)/2}\simeq -V^r$ 
while $V^y\simeq0$. Hence the observer is accelerated 
to almost a null radial geodesic as she
nears $r=0$ and $L$ approaches an infinite radial boost.
In this limit, one may drop the $V^y$ terms in eq.~\reef{loren}.

With four $L$'s in the transformation of the curvature,
naively the worst divergence would be $O((V^r)^4)$, which would appear
in $\hat R^{trtr}$. However, one finds using 
$(V^t)^2-(V^r)^2=1$ that $\hat R^{trtr}=R^{trtr}$.
So as a result of a precise cancellation of terms,
only a finite contribution remains.
Similarly for components where three $L$'s appear, one finds
\beq
\hat R^{tatr} 
\simeq |V^r|(R^{tatr}-R^{ratr})\simeq-\hat R^{ratr}
\label{yayab}
\eeq
where $a\ne r,t$.
These components diverge but only as a single power of $V^r$,
and in fact, these singularities would vanish in the special case that
$R^{ratr}=R^{tatr}$.
The previous 
divergences are in fact subleading compared to components such as
\beqa
\hat R^{tatb} 
&=&(V^r)^2(R^{tatb}-R^{tarb}-R^{ratb}+R^{rarb})+ R^{tatb}
\labell{yayac}\\
\hat R^{rarb} &=& (V^r)^2(R^{tatb}-R^{tarb}-R^{ratb}+R^{rarb})+R^{rarb}
\nonumber\\
\hat R^{tarb} &=& -(V^r)^2(R^{tatb}-R^{tarb}-R^{ratb}+R^{rarb})+R^{tarb}
\nonumber
\eeqa
where $a,b\ne r,t$. So here one finds the naively expected
divergence of $O((V^r)^2)$, except for the exceptional circumstance
that $R^{tatb}-R^{tarb}-R^{ratb}+R^{rarb}=0$, in which case
these components are invariant.
Thus the geometric symmetries of the
Riemann tensor dictate that the worst divergence is only $(V^r)^2$.

Implementing the transformation \reef{loren} to the
curvature components \reef{curvapp}, the
leading divergences appear in 
\begin{eqnarray}
\hat{R}^{tyty}&\simeq& -{\beta(\beta-1)B_0\over2Q}(\omega-p)^2\, r^{(-\beta-1)}
\labell{wurst}\\
\hat{R}^{ta tb}&\simeq&\delta^{ab}
{l(l+1)B_0\over4Q}(\omega-p)^2\, r^{(-\beta-1)}\qquad\ {\rm for}\ a,b=\theta,
\phi
\nonumber\\
\hat{R}^{titj}&\simeq&\delta^{ij}{n_i^2\over 2R_i^2}B_0Q(\omega-p)^2\,
r^{(-\beta-1)}\qquad\quad{\rm for}\ i,j=5\ldots9
\nonumber
\eeqa
as well as $\hat{R}^{rarb}\simeq\hat{R}^{tatb}\simeq-\hat{R}^{tarb}$.
Hence quite generically the perturbations \reef{constprof} produce
singular tidal forces on the null surface $r=0$.
Because these divergences will not be cancelled by any
other terms of the metric for slowly oscillating waves, one may
conclude that all of the new solutions have a null singularity at $r=0$.
Hence, the excitation of the wave-like perturbations
generically destroys the horizon, producing a
naked singularity instead.
There is one exception to this conclusion. If $l=0$ and
$n_i=0$, for which $\beta=1$, 
the coefficients of the potentially divergent terms all vanish
--- this applies to both the quadratic and linear divergences. 
In this special case, the remaining curvature components \reef{curvapp}
are boost invariant. Hence it seems that there
is a single family of wavy perturbations which qualifies as
black hole hair.

\section{DISCUSSION\labels{sec:end}}

We have shown that the wave-generating technique of Garfinkle and
Vachaspati can be applied to certain higher dimensional black holes.
While all scalar curvature invariants are left unchanged by this
construction, generically the new waves produce
parallelly propagated curvature singularities.
Hence in these cases, the horizon is destroyed and 
a null singularity is produced instead. Thus it seems that
these higher dimensional black holes still respect the
spirit of the no-hair theorems. The only nonsingular waves
are those with $l=0=n_i$. In this case, the original solution \reef{sol11}
is mapped to
\beqa
ds^2&=&- {\left(1-{Q\over R}\right)^2\over 1+{b(u)\over R}} dt^2+
{dR^2\over\left(1-{Q\over R}\right)^2}+ R^2(d\theta^2
+\sin^2\theta\,d\phi^2)
\nonumber\\
\labell{sol12}
&&\qquad\qquad+\left(1+{b(u)\over R}\right)\left(dy-{Q+b(u)\over R+b(u)}
dt\right)^2+\sum_{i=5}^9 (dx^i)^2
\eeqa
where the wave profile $b(u)$ has an arbitrary dependence on $u=t-y$.
A constant $b$ would yield a shift in the momentum flowing along $y$.
In general, these perturbations represent longitudinal waves
carrying momentum in the $y$ direction without transverse oscillations.
These nonsingular waves are also distinguished in that they are the
only localized waves which drop off slowly enough, \ie $1/R$, to be detected
in the asymptotic region in either the energy or momentum density\cite{us}.

While no evidence of a curvature singularity has been found in the
present analysis, it may be that more subtle singularities remain
to be uncovered --- see for example \cite{welch}.
Normally the approach to proving the existence of
a singularity-free horizon would be to find coordinates in which the metric
is analytic at the null surface in question. Some progress has been
made in finding such coordinates\cite{us}. An essential feature
of the examples where such coordinates have been found is that $b(u)\rightarrow
constant$ as $u\rightarrow\pm\infty$. In fact, when the profile
does not approach a constant, Horowitz and Yang\cite{HY} have found that
a mild singularity is produced. In this case, the nonsingular perturbations
seem not to represent wavy hair for these black holes, but rather
transient waves similar to those which might be produced in a
gravitational collapse producing a black hole.

From this point of view, it may seem suprising that the remaining
localized waves produced singularities, since one might have expected
that a gravitational collapse could also produce such waves.
One can investigate
the strength of the singularities for $\beta>1$. Along the geodesics
investigated above, one finds that the divergent curvatures
$\hat{R}\propto \tau^{-2}$, where $\tau$ is the proper time along
the curve. Hence these singularities are by no means integrable, and
would result in an extended probe being crushed.

As a final comment, we should mention that in ref.~\cite{andy},
string theory was used to produce a statistical mechanical understanding
of the black hole entropy for the solution in eq.~\reef{sol11}.
In ref.~\cite{HM}, these calculations were
extended to include certain wavy excitations, including that
in eq.~\reef{sol12}.
It is interesting that the latter analysis succeeded despite the
presence of the singularity later found in ref.~\cite{HY}.

\section*{ACKNOWLEDGMENTS}

\indent I would like to thank Nemanja Kaloper and Harold Roussel for
a stimulating collaboration.
This work was supported in part by NSERC of Canada and in
part by Fonds FCAR du Qu\'ebec.

\end{document}